\documentstyle [12pt] {article}

\catcode`@=12
\begin{document}
\thispagestyle{empty}
\begin{flushright} UCRHEP-T159\\August 1996\
\end{flushright}
\vspace{0.5in}
\begin{center}
{\Large	\bf Tree-Level Nondecoupling and\\}
\vspace{0.1in}
{\Large \bf the Supersymmetric Higgs Sector\\}
\vspace{1.0in}
{\bf Xiao-yuan Li\\}
\vspace{0.1in}
{\sl Institute of Theoretical Physics, Chinese Academy of Sciences\\
P.~O.~Box 2735, Beijing 100080, People's Republic of China\\}
\vspace{0.2in}
{\bf Ernest Ma\\}
\vspace{0.1in}
{\sl Department of Physics, University of California\\
Riverside, California 92521, U.~S.~A.\\}
\vspace{0.9in}
\end{center}

\begin{abstract}\
Because of the existence of cubic scalar couplings, there are in general 
nondecoupling effects at tree level in the scalar sector of any theory 
with two or more very different mass scales.  We show this explicitly in 
the minimal nonsupersymmetric SU(5) model of grand unification.  We show 
also how tree-level decoupling is guaranteed if supersymmetry is imposed. 
On the other hand, if the gauge symmetry is larger than that of the 
standard model at the mass scale of supersymmetry breaking, the 
two-Higgs-doublet structure at the presumably lower electroweak energy 
scale will be different from that of the minimal 
supersymmetric standard model, as shown already previously in a number of 
specific examples.  We add here one example involving four Higgs doublets.
\end{abstract}

\newpage
\baselineskip 24pt

\section{Introduction}

In the study of fundamental interactions, it is important to recognize the 
relevant energy scale or scales of the specific processes being discussed. 
In quantum field theory, it means knowing whether an interaction at a 
particular energy scale has an inseparable component from physics at a 
much higher scale.  If not, then there is decoupling, and the theoretical 
interpretation of any experimental result becomes tractable in the 
context of that theory.  On the other hand, if there is nondecoupling, 
then an irreducible degree of uncertainty must always remain.

In the standard electroweak gauge model, although there is really only one 
energy scale, {\it i.e.} $v = (2 \sqrt 2 G_F)^{-1/2} \simeq 170$ GeV, 
nondecoupling is 
known to occur in the limit of large fermion masses.  We review this 
in Sec.~2 and show that in addition to the commonly touted loop effects, 
nondecoupling is already present at tree level.  We then discuss in Sec.~3 
the case of two very different energy scales and show how nondecoupling 
occurs at tree level in the Higgs sector with two explicit examples: one of 
$SU(5)$ breaking down to $SU(3) \times SU(2) \times U(1)$, and the other of 
two scalar doublets in the standard model.  In Sec.~4 we 
show how exact supersymmetry guarantees the decoupling of the two scales.  
In Sec.~5, we show how softly broken supersymmetry allows nondecoupling 
and apply it to the case where the gauge symmetry is larger than that of 
the standard model at the mass scale of supersymmetry breaking.  As shown 
in several previous specific examples\cite{1,2,3}, the two-Higgs-doublet 
structure at the presumably lower electroweak energy scale will be 
different from that of the minimal supersymmetric standard model (MSSM).  
We add here one example involving four Higgs doublets.  Finally in Sec.~6, 
there are some concluding remarks.

\section{Nondecoupling in the Standard Model}

The decoupling\cite{4} of particles heavier than a certain mass scale from 
physics at a much lower energy is important for the proper interpretation of 
experimental observables in terms of a particular theory.  In the standard 
electroweak gauge model, there is technically only one scale, {\it i.e.} 
the vacuum expectation value of the neutral component of the Higgs scalar 
doublet $v = (2 \sqrt 2 G_F)^{-1/2} \simeq 174$ GeV.  Whereas all masses 
in this model are proportional to $v$, it is still meaningful to consider the 
limit that one of these masses is much larger than all the others, and ask 
if the former's contribution to physically measurable quantities vanishes 
or not.  It has been known for some time that nondecoupling of heavy 
particles does occur\cite{5} in models with spontaneous symmetry breaking, 
and since the heaviness of the $t$ quark is now established, {\it i.e.} 
$m_t = 180 \pm 12$ GeV\cite{6}, its contributions to many observables 
are confirmed to be nonvanishing and nonnegligible.  The lesson we learn 
here is that without knowing the value of $m_t$, there would be large 
uncertainties in the interpretation of data in terms of the standard model.

Examples of nondecoupling in the standard model abound, but they have been 
invariably given as loop effects.  Consider the process $H \rightarrow 
\gamma \gamma$, where $H$ is the standard-model Higgs boson.  Since the 
Yukawa coupling of $H$ to $\bar t t$ is proportional to $m_t/v$, the $t$ 
contribution to this amplitude is of the form
\begin{equation}
{m_t \over v} ~ \times ~ {e^2 \over {16 \pi^2}} {1 \over m_t},
\end{equation}
which goes to a nonzero constant as $m_t \rightarrow \infty$.  In other words, 
the suppression of the loop due to a large $m_t$ is exactly compensated 
by the increased coupling.  If $t$ is replaced with the $W$ boson, then 
we have instead
\begin{equation}
g^2 v ~ \times ~ {e^2 \over {16 \pi^2}} {1 \over {g^2 v^2}},
\end{equation}
which is again finite as $M_W = gv/\sqrt 2$ goes to infinity, because we 
must hold $v$ finite to get a finite $m_H$.  In the above, 
there is always the implicit but crucial assumption that $m_H << m_t$ and 
$m_H << M_W$, because we are concerned with the effect of heavy particles 
on experimental observables far below the energy required to produce these 
heavy particles.

Another example is the one-loop radiative correction to the $W$ and $Z$ 
self-energies.  The oblique parameter $T$\cite{7} is proportional to 
the difference between $\Pi_{11}(0) = \Pi_{22}(0)$ and $\Pi_{33}(0)$, which 
has a contribution from the $t$ and $b$ quarks of the form
\begin{equation}
{g^2 \over {16 \pi^2}} \left[ m_t^2 + m_b^2 - {{2 m_t^2 m_b^2 \ln 
(m_t^2/m_b^2)} \over {m_t^2 - m_b^2}} \right],
\end{equation}
which is always nonnegative and is zero only if $m_t = m_b$.  It also 
increases without bound as $m_t$ (or $m_b$) increases.  The implicit 
assumption here is that the $W$ and $Z$ masses are fixed.

Actually, nondecoupling in the standard model already occurs at tree level. 
Consider the interaction of four light particles.  Divide them into two pairs. 
If a heavy particle (of mass $M$) couples to each pair, then an effective 
coupling of the form $g_1 g_2/M^2$ appears.  In the standard model, in 
the case of four light fermions, $g_1 = g_2 = g/2 \sqrt 2$ and 
$M^2 = g^2 v^2/2$, hence the effective 
coupling is $1/4 v^2$ which does not vanish as $M \rightarrow \infty$. 
This is just like the previous example regarding $T$, except now the 
Yukawa coupling $f$ in $m_f = fv$ is assumed small compared to the 
gauge coupling $g$ instead of the other way around.

\section{Nondecoupling in the Case of Two Scales}

Consider now a gauge group larger than that of the standard model.  Let the 
former break down to the latter at the scale $v_H$ which is much larger 
than the electroweak scale $v$.  Let $H$ be a heavy scalar boson (with mass 
proportional to $v_H$) which is a singlet under the standard $SU(3) \times 
SU(2) \times U(1)$ gauge symmetry.  Let $\Phi$ be the usual standard Higgs 
doublet and assume that the cubic interaction $\Phi^\dagger \Phi H$ exists. 
Now if this coupling strength is proportional to $v_H$ also, then the 
effective $(\Phi^\dagger \Phi)^2$ coupling is of the form
\begin{equation}
v_H ~ {1 \over v_H^2} ~ v_H,
\end{equation}
which does not vanish as $v_H \rightarrow \infty$.  Here the nondecoupling 
of the heavy particle $H$ has occurred at tree level in spite of the large 
ratio $v_H/v$, in contrast to the last example in the previous section where 
the ratio $g/f$ is large but there is only one scale.

In a nonsupersymmetric quantum field theory with two (or more) scales, 
the above nondecoupling phenomenon in the scalar sector is a general 
occurrence.  In the next section, we will show how an exactly supersymmetric 
theory enforces the decoupling of the two scales.  Here we provide first a 
very useful example of nonsupersymmetric $SU(5)$ breaking into $SU(3) \times 
SU(2) \times U(1)$ in the presence of one adjoint {\bf 24} and one 
fundamental {\bf 5} of Higgs scalar representations.  It is often 
implicitly assumed here that electroweak symmetry breaking is determined by 
the quartic self-interaction of the doublet scalar field $\Phi$ in the 
$\bf 5$ which appears in the $SU(5)$ Lagrangian.  However, it will be shown 
in the following that because of nondecoupling contributions from heavy 
particles contained in the {\bf 24}, such is not the case.  One important 
consequence of this result is that if one uses the renormalization group 
equations to run the former coupling from the $SU(5)$ scale to the 
electroweak scale, it will not be the experimentally observed coupling.

Let the adjoint {\bf 24} scalar representation be denoted by a $5 \times 5$ 
matrix:
\begin{equation}
H = \left[ \begin{array} {c@{\quad}c} H_{\alpha \beta} - (2/15)^{1/2} H_0 
\delta_{\alpha \beta} & H_{\alpha j} \\ H_{i \beta} & H_{ij} + (3/10)^{1/2} 
H_0 \delta_{ij} \end{array} \right],
\end{equation}
where $\alpha, \beta = 1,2,3$; $i,j = 4,5$; and
\begin{equation}
H_{ij} = 2^{-1/2} \vec \tau \cdot \vec H_3, ~~~ H_{\alpha \beta} = 2^{-1/2} 
\vec \lambda \cdot \vec H_8.
\end{equation}
In the above, $\vec \tau$ denotes the 3 $SU(2)$ $2 \times 2$ representation 
matrices and $\vec \lambda$ the 8 $SU(3)$ $3 \times 3$ ones.  The vacuum 
expectation value of $H$ is assumed to be given by
\begin{equation}
\langle H \rangle = {{\langle H_0 \rangle} \over {\sqrt {30}}} \left[ 
\begin{array} {c@{\quad}c@{\quad}c@{\quad}c@{\quad}c} -2 & & & & \\ 
 & -2 & & & \\ & & -2 & & \\ & & & 3 & \\ & & & & 3 \end{array} \right] + 
{{\langle H_3^0 \rangle} \over {\sqrt 2}} \left[ \begin{array} 
{c@{\quad}c@{\quad}c@{\quad}c@{\quad}c} 0 & & & & \\ & 0 & & & \\ 
& & 0 & & \\ & & & 1 & \\ & & & & -1 \end{array} \right].
\end{equation}
Note that the usual discussion of $SU(5)$ symmetry breaking routinely 
neglects the triplet scalar field $\vec H_3$ as well as its vacuum 
expectation value.  However, it will be shown in the following that they 
are essential in 
correctly understanding the symmetry breaking.  The fundamental {\bf 5} 
is denoted by $\Phi = [ \Phi_\alpha, \Phi_i ]$, with $\Phi_i = (\phi^+, 
\phi^0)$; hence $\langle \Phi \rangle = \langle \phi^0 \rangle 
[0,0,0,0,1]$.

The most general Higgs potential consisting of $H$ and $\Phi$ which is also 
invariant under the discrete symmetry $H \rightarrow -H$ is given by
\begin{eqnarray}
V &=& {1 \over 2} m_1^2 Tr H^2 + {1 \over 4} \lambda_1 (Tr H^2)^2 + 
{1 \over 4} \lambda_2 Tr H^4 \nonumber \\ &+& m_2^2 \Phi^\dagger \Phi + 
{1 \over 2} \lambda_3 (\Phi^\dagger \Phi)^2 + \lambda_4 (Tr H^2)(\Phi^\dagger 
\Phi) + \lambda_5 (\Phi^\dagger H^2 \Phi).
\end{eqnarray}
Let
\begin{equation}
\langle H_0 \rangle = v_1, ~~~ \langle \phi^0 \rangle = v_2/\sqrt 2, ~~~ 
\langle H_3^0 \rangle = v_3,
\end{equation}
then the minimum of $V$ satisfies
\begin{eqnarray}
&~& v_1 [ m_1^2 + (\lambda_1 + {7 \over 30} \lambda_2) v_1^2 + (\lambda_4 + 
{3 \over 10} \lambda_5) v_2^2 + (\lambda_1 + {9 \over 10} \lambda_2) v_3^2 ] 
-{1 \over 2} \sqrt {3 \over 5} \lambda_5 v_2^2 v_3 = 0, \\ 
&~& v_2 [ m_2^2 + (\lambda_4 + {3 \over 10} \lambda_5) v_1^2 + {1 \over 2} 
\lambda_3 v_2^2 - \sqrt {3 \over 5} \lambda_5 v_1 v_3 + {1 \over 2} 
\lambda_5 v_3^2 ] = 0, \\ 
&~& v_3 [ m_1^2 + (\lambda_1 + {9 \over 10} \lambda_2) v_1^2 + (\lambda_4 + 
{1 \over 2} \lambda_5) v_2^2 + (\lambda_1 + {1 \over 2} \lambda_2) v_3^2 ] 
-{1 \over 2} \sqrt {3 \over 5} \lambda_5 v_2^2 v_1 = 0.
\end{eqnarray}
Note that $v_1 \neq 0, v_2 = v_3 = 0$ is a solution; but if $v_2 \neq 0$ 
as well, then $v_3 \neq 0$ necessarily.  The often quoted naive solution 
without $v_3$ is not strictly correct, but since its magnitude is of order 
$v_2^2/v_1$, it is negligible for all known purposes.

Solving the above 3 equations, using the valid approximation that 
$v_1 >> v_2 >> v_3$, we obtain
\begin{equation}
v_1^2 \simeq {{-m_1^2} \over {\lambda_1 + 7 \lambda_2/30}},
\end{equation}
and
\begin{equation}
v_2^2 \simeq {{-m_2^2 + (\lambda_4 + 3 \lambda_5/10) m_1^2/(\lambda_1 + 
7 \lambda_2/30)} \over {\lambda_3/2 - 9 \lambda_5^2/ 20 \lambda_2 - 
(\lambda_4 + 3 \lambda_5/10)^2/(\lambda_1 + 7 \lambda_2/30)}}.
\end{equation}
Let us go back to $V$ and consider the $\Phi_i^\dagger \Phi_i$ term.  This 
is correctly given in the usual treatment as
\begin{equation}
\mu^2 = m_2^2 + \lambda_4 v_1^2 + {3 \over 10} \lambda_5 v_1^2,
\end{equation}
but then it is often claimed that
\begin{equation}
v_2^2 = {{-\mu^2} \over {\lambda_3/2}} = {{-m_2^2 + (\lambda_4 + 
3 \lambda_5/10) m_1^2/(\lambda_1 + 7 \lambda_2/30)} \over {\lambda_3/2}},
\end{equation}
which is of course wrong.  The two missing terms in the denominator are 
exactly those given by the nondecoupling contributions of $H_0$ and 
$\vec H_3$.  It is easy to verify that $H_0$ has mass-squared = $2(\lambda_1 
+ 7 \lambda_2/30) v_1^2$ and its coupling to $\Phi_i^\dagger \Phi_i$ is 
$2(\lambda_4 + 3 \lambda_5/10) v_1$; and $\vec H_3$ has mass-squared = 
$(2 \lambda_2/3) v_1^2$ and its coupling to $\Phi_i^\dagger \vec \tau_{ij} 
\Phi_j$ is $\sqrt {3/5} \lambda_5 v_1$.  Hence the correct quartic 
self-coupling of the Higgs doublet $\Phi$ at low energy is
\begin{equation}
\lambda_3 - {9 \over 10} {\lambda_5^2 \over \lambda_2} - {{2(\lambda_4 + 
3 \lambda_5/10)^2} \over {\lambda_1 + 7 \lambda_2/30}}.
\end{equation}
In other words, there are inseparable contributions from heavy particles 
at the large scale $v_1$ to the low-energy interactions of light particles 
at the small scale $v_2$.  The only way that these contributions can be 
discovered is to increase the experimental energy scale up to $v_1$.

Another demonstration of this kind of tree-level nondecoupling is available 
in the well-known extension of the standard model to include two Higgs 
doublets.  Let the Higgs potential be given by
\begin{eqnarray}
V &=& \mu_1^2 \Phi_1^\dagger \Phi_1 + \mu_2^2 \Phi_2^\dagger \Phi_2 + 
{1 \over 2} \lambda_1 (\Phi_1^\dagger \Phi_1)^2 + {1 \over 2 } \lambda_2 
(\Phi_2^\dagger \Phi_2)^2 \nonumber \\ &+& \lambda_3 (\Phi_1^\dagger \Phi_1) 
(\Phi_2^\dagger \Phi_2) + \lambda_4 (\Phi_1^\dagger \Phi_2) ( \Phi_2^\dagger 
\Phi_1) + {1 \over 2} \lambda_5 [(\Phi_1^\dagger \Phi_2)^2 + 
(\Phi_2^\dagger \Phi_1)^2].
\end{eqnarray}
Let $\langle \phi^0_{1,2} \rangle = v_{1,2}$, then
\begin{eqnarray}
v_1 [ \mu_1^2 + \lambda_1 v_1^2 + (\lambda_3 + \lambda_4 + \lambda_5) v_2^2 ] 
&=& 0, \\ v_2 [ \mu_2^2 + \lambda_2 v_2^2 + (\lambda_3 + \lambda_4 + 
\lambda_5) v_1^2 ] &=& 0.
\end{eqnarray}
Suppose $v_2 << v_1$, then $v_1^2 \simeq - \mu_1^2/\lambda_1$ and 
\begin{equation}
v_2^2 \simeq {{-\mu_2^2 + (\lambda_3 + \lambda_4 + \lambda_5) \mu_1^2/
\lambda_1} \over {\lambda_2 - (\lambda_3 + \lambda_4 + \lambda_5)^2/ 
\lambda_1}}.
\end{equation}
Again, the heavy $\phi_1^0$ contribution is nondecoupling.  An important 
note is that the two scales $v_1$ and $v_2$ can be separated in principle 
here because $V$ has a discrete symmetry $\Phi_1 \rightarrow \Phi_1$, 
$\Phi_2 \rightarrow -\Phi_2$, which is unbroken in the case $v_1 \neq 0$, 
$v_2 = 0$.

\section{Guarantee of Decoupling in Supersymmetry}

In a supersymmetric quantum field theory, the Higgs potential $V$ is 
necessarily nonnegative.  However, there may be several minima, each 
having $V=0$ but corresponding to different sets of vacuum expectation values. 
Hence the gauge symmetry may be broken while the supersymmetry is preserved. 
For example in the case supersymmetric $SU(5)$, it has been shown\cite{8} 
that with an adjoint {\bf 24}, the supersymmetric-preserving vacuum may be 
symmetric under $SU(5)$ ({\it i.e.} no breaking), $SU(4) \times U(1)$, or 
$SU(3) \times SU(2) \times U(1)$ ({\it i.e.} the standard model). [In the 
notation of the previous section, these solutions correspond to $v_1 = v_3 
= 0$; $v_1 = 0$, $v_3 \neq 0$; and $v_1 \neq 0$, $v_3 = 0$ respectively.] 
To understand how exact supersymmetry guarantees the decoupling of heavy 
particles from low-energy physics, we need only consider the structure of 
$V$ given by the superpotential $W$, as follows.

To be specific, consider the two usual doublet superfields of the 
supersymmetric standard model:
\begin{equation}
\tilde \Phi_1 \equiv i \tau_2 \Phi_1^* = \left( \begin{array} {c} 
\bar \phi_1^0 \\ - \phi_1^- 
\end{array} \right), ~~~ \Phi_2 = \left( \begin{array} {c} \phi_2^+ \\ 
\phi_2^0 \end{array} \right).
\end{equation}
Introduce the heavy superfields $S$ and $\vec \Sigma$ (singlet and triplet 
respectively under the standard electroweak gauge group).  Then the 
relevant superpotential involving the above is given by
\begin{equation}
W = {1 \over 2} m_S S S + {1 \over 2} m_\Sigma \vec \Sigma \cdot \vec \Sigma 
+ \mu \Phi_1^\dagger \Phi_2 + f_S S \Phi_1^\dagger \Phi_2 + {1 \over 2} 
f_\Sigma \Phi_1^\dagger (\vec \Sigma \cdot \vec \tau) \Phi_2,
\end{equation}
where $\mu << m_S, m_\Sigma$.  If supersymmetry is exact, then the part of 
the Higgs potential which comes from $W$ is given by
\begin{eqnarray}
V_F &=& |m_S S + f_S \Phi_1^\dagger \Phi_2|^2 + |m_\Sigma \vec \Sigma + 
{1 \over 2} f_\Sigma \Phi_1^\dagger \vec \tau \Phi_2|^2 \nonumber \\ 
&+& |f_S S \Phi_2 + {1 \over 2} f_\Sigma \vec \Sigma \cdot \vec \tau \Phi_2 
+ \mu \Phi_2|^2 + | f_S \Phi_1^\dagger S + {1 \over 2} f_\Sigma \Phi_1^\dagger 
\vec \Sigma \cdot \vec \tau + \mu \Phi_1^\dagger|^2,
\end{eqnarray}
which is clearly nonnegative.  Now the cubic $\Phi^\dagger_{1,2} \Phi_{1,2} S$ 
coupling strength is $f_S \mu$, hence the $S$ contribution to 
$(\Phi^\dagger_{1,2} \Phi_{1,2})^2$ is of the form
\begin{equation}
f_S \mu ~ {1 \over m_S^2} ~ f_S \mu,
\end{equation}
which goes to zero as $m_S/\mu \rightarrow \infty$.  On the other hand, the 
cubic $\Phi_1^\dagger \Phi_2 S$ coupling strength is $f_S m_S$ which is 
large, but $V_F$ also contains an explicit $|\Phi_1^\dagger \Phi_2|^2$ term, 
hence the $S$ contribution here is given by
\begin{equation}
f_S^2 - f_S m_S ~ {1 \over m_S^2} ~ f_S m_S,
\end{equation}
which is zero.  Similarly, the $\vec \Sigma$ contributions also decouple. 
In fact, the only term which survives is 
$\mu^2 (\Phi_1^\dagger \Phi_1 + \Phi_2^\dagger \Phi_2)$.  There is also 
a part of $V$ which comes from the gauge sector but it has no cubic 
interactions and thus no additional tree-level contributions from the 
heavy scalar fields.  In the notation of Eq.~(18), the quartic scalar 
couplings are given by
\begin{equation}
\lambda_1 = \lambda_2 = {1 \over 4} (g_1^2 + g_2^2), ~~ \lambda_3 = 
-{1 \over 4} g_1^2 + {1 \over 4} g_2^2, ~~ \lambda_4 = -{1 \over 2} g_2^2, 
~~ \lambda_5 = 0.
\end{equation}
This means that as long as supersymmetry is maintained exactly above the 
electroweak symmetry breaking scale, the two-Higgs-doublet structure is 
uniquely given by the minimal supersymmetric standard model (MSSM).  
However, the scale of soft supersymmetry breaking $M_{SUSY}$ may be 
somewhat higher, say a few TeV instead of 100 GeV, so there is room enough 
for a larger gauge symmetry to be in effect above $M_{SUSY}$.  It 
must of course also break down to the standard $SU(3) \times SU(2) \times 
U(1)$ below $M_{SUSY}$.  In this case, the nondecoupling of heavy particles 
at the $M_{SUSY}$ scale will change the quartic scalar couplings listed 
above, as already shown in several previous explicit examples\cite{1,2,3}.

\section{Nondecoupling with Softly Broken Supersymmetry}

Consider the following scenario of symmetry breaking:
\vspace{0.2in}

\begin{tabular} {|c|c|c|} \hline
energy & gauge group & supersymmetry \\ \hline
$10^{16}$ GeV & $G \rightarrow G'$ & unbroken \\
$10^x$ GeV & $G' \rightarrow G''$ & unbroken \\ \hline
$10^3$ GeV & $G'' \rightarrow G_{SM}$ & broken \\
$10^2$ GeV & $G_{SM} \rightarrow SU(3) \times U(1)$ & broken \\ \hline
\end{tabular}
\vspace{0.2in}

In the above, $G_{SM}$ is the standard-model gauge group and $10^x$ GeV 
is a possible but unknown intermediate scale.  The usual assumption is 
that $G'' = G_{SM}$ in which case supersymmetry would protect the MSSM 
up to $10^x$ GeV.  Often it is also assumed that $G' = G''$, in which 
case $x = 16$.  However, if $G''$ is larger than $G_{SM}$, then the 
physics at $10^2$ GeV will have nondecoupling contributions from $G''$.

In addition to previous examples\cite{1,2,3} of two Higgs doublets at the 
electroweak scale which are not those of the MSSM, consider here a model 
which ends up with four Higgs doublets.  It is the supersymmetric version 
of a gauge model of generation nonuniversality proposed many years 
ago\cite{9}.  We extend the electroweak gauge group to $SU(2)_{12} \times 
SU(2)_3 \times U(1)$ with couplings $g_{12}$, $g_3$, and $g_0$, such that 
left-handed quark and lepton doublets of the first two generations couple 
to $SU(2)_{12}$ but those of the third couple to $SU(2)_3$.  To make 
this into a supersymmetric theory, in analogy to the doubling of Higgs 
scalars in the standard model, we need the following scalar multiplets:
\begin{eqnarray}
\Phi_1 = \left( \begin{array} {c} \phi_1^+ \\ \phi_1^0 \end{array} \right) 
\sim (2, 1, {1 \over 2}), &~& \Phi_2 = \left( \begin{array} {c} \bar \phi_2^0 
\\ -\phi_2^- \end{array} \right) \sim (2, 1, -{1 \over 2}), \\ 
\Phi_3 = \left( \begin{array} {c} \phi_3^+ \\ \phi_3^0 \end{array} \right) 
\sim (1, 2, {1 \over 2}), &~& \Phi_4 = \left( \begin{array} {c} \bar \phi_4^0 
\\ -\phi_4^- \end{array} \right) \sim (1, 2, -{1 \over 2}),
\end{eqnarray}
\begin{equation}
\eta = \left( \begin{array} {c@{\quad}c} \bar \eta_2^0 & \eta_1^+ \\ -\eta_2^- 
& \eta_1^0 \end{array} \right) \sim (2, 2, 0), ~~~ S \sim (1, 1, 0).
\end{equation}
The spontaneous breaking of $SU(2)_{12} \times SU(2)_3$ to $SU(2)_{SM}$ is 
achieved by having $\langle \eta_1^0 \rangle = \langle \bar \eta_2^0 
\rangle = u$.  The singlet $S$ is added in the above because a careful 
examination of the Higgs potential shows that without it, there would be 
no such solution.  From the gauge interactions alone, the Higgs potential 
is given by
\begin{eqnarray}
V_D &=& {1 \over 8} g_0^2 (\Phi_1^\dagger \Phi_1 - \Phi_2^\dagger \Phi_2 + 
\Phi_3^\dagger \Phi_3 - \Phi_4^\dagger \Phi_4)^2 \nonumber \\ &+& 
{1 \over 8} g_{12}^2 \sum_a (\Phi_1^\dagger \tau^a \Phi_1 + \Phi_2^\dagger 
\tau^a \Phi_2 + Tr \eta^\dagger \tau^a \eta)^2 \nonumber \\ &+& 
{1 \over 8} g_3^2 \sum_a (\Phi_3^\dagger \tau^a \Phi_3 + \Phi_4^\dagger 
\tau^a \Phi_4 - Tr \eta \tau^a \eta^\dagger)^2.
\end{eqnarray}
From the superpotential
\begin{eqnarray}
W &=& \mu_1 \tilde \Phi_1^\dagger \Phi_2 + \mu_2 \tilde \Phi_3^\dagger \Phi_4 
+ \mu_3 Tr \tilde \eta^\dagger \eta + {1 \over 2} \mu_4 S^2 + 
f_1 \tilde \Phi_1^\dagger \eta \Phi_4 
+ f_2 \tilde \Phi_2^\dagger \eta \Phi_3 \nonumber \\ &+& \lambda_1 S \tilde 
\Phi_1^\dagger \Phi_2 + \lambda_2 S \tilde \Phi_3^\dagger \Phi_4 + \lambda_3 
S Tr \tilde \eta^\dagger \eta + {1 \over 3} \lambda_4 S^3,
\end{eqnarray}
we obtain
\begin{eqnarray}
V_F &=& |\lambda_1 \tilde \Phi_1^\dagger \Phi_2 + \lambda_2 \tilde 
\Phi_3^\dagger \Phi_4 + \lambda_3 Tr \tilde \eta^\dagger \eta + 
\lambda_4 S^2 + \mu_4 S|^2 \nonumber \\ &+& \sum_{i,j} |2 \mu_3 \tilde 
\eta_{ji}^\dagger + 2 \lambda_3 S \tilde \eta_{ji}^\dagger + f_1 \tilde 
\Phi_{1i}^\dagger \Phi_{4j} + f_2 \tilde \Phi_{2i}^\dagger \Phi_{3j}|^2 
\nonumber \\ &+& |\mu_1 \Phi_2 + \lambda_1 S \Phi_2 + f_1 \eta \Phi_4|^2 + 
|-\mu_1 \Phi_1 - \lambda_1 S \Phi_1 + f_2 \eta \Phi_3|^2 \nonumber \\ 
&+& |-\mu_2 \tilde \Phi_4^\dagger - \lambda_2 S \tilde \Phi_4^\dagger + 
f_2 \tilde \Phi_2^\dagger \eta|^2 + |\mu_2 \tilde \Phi_3^\dagger + \lambda_2 
S \tilde \Phi_3^\dagger + f_1 \tilde \Phi_1^\dagger \eta|^2,
\end{eqnarray}
where $\tilde \eta \equiv \tau_2 \eta^* \tau_2$, and $\tilde \Phi_1^\dagger 
\Phi_2 = -\tilde \Phi_2^\dagger \Phi_1$ has been used.

As $\eta$ and $S$ acquire vacuum expectation values $u_1 = u_2 = u$ and $s$ 
respectively, this model reduces to the standard model as far as the 
gauge interactions are concerned.  All left-handed quarks and leptons are 
now doublets under $SU(2)_{SM}$ with coupling strength $g_{123}$ given 
by\cite{9}
\begin{equation}
g_{123}^{-2} = g_{12}^{-2} + g_3^{-2}.
\end{equation}
In the Higgs sector, the 8 scalar fields contained in the bidoublet $\eta$ 
are now organized into a massless triplet ({\it i.e.} the would-be 
Goldstone bosons of this symmetry breaking), a massive triplet [$Re(\eta_1^0 
- \eta_2^0), (\eta_1^\pm - \eta_2^\pm)/\sqrt 2$], and two singlets, {\it i.e.} 
$Re(\eta_1^0 + \eta_2^0)$ and $Im(\eta_1^0 - \eta_2^0)$, the first of which 
also mixes with $ReS$.  Assume for simplicity
\begin{equation}
V_{soft} = \mu^2 Tr \eta^\dagger \eta + m^2 |S|^2,
\end{equation}
then the triplet [$Re(\eta_1^0 - \eta_2^0), (\eta_1^\pm - \eta_2^\pm)/
\sqrt 2$] has mass-squared given by
\begin{equation}
M^2 = (g_{12}^2 + g_3^2) u^2 - 4 \lambda_3 (2 \lambda_3 u^2 + \lambda_4 s^2 
+ \mu_4 s),
\end{equation}
and couples to $\Phi_{1,2}^\dagger \vec \tau \Phi_{1,2}$ with strength 
$-(1/2) g_{12}^2 u$, and to $\Phi_{3,4}^\dagger \vec \tau \Phi_{3,4}$ with 
strength $(1/2) g_3^2 u$.  The effective $(\Phi_{1,2}^\dagger \Phi_{1,2})^2$ 
interaction is thus
\begin{equation}
{1 \over 2} g_{12}^2 - {{2(g_{12}^2 u/2)^2} \over M^2},
\end{equation}
which reduces to $g_{123}^2/2$ if we drop terms in $M^2$ having to do with the 
superpotential.  The same result holds for the effective $(\Phi_{3,4}^\dagger 
\Phi_{3,4})^2$ interaction with the interchange of $g_{12}^2$ and $g_3^2$.  
Other nondecoupling contributions also appear, 
but their cubic interactions do not involve the gauge couplings.  The 
four-doublet structure of the reduced Higgs potential has thus many 
parameters and is not simply a function of the standard-model gauge 
couplings.

\section{Concluding Remarks}

We have demonstrated in this paper that tree-level nondecoupling occurs 
generally in the scalar sector of a spontaneously broken gauge theory. 
The origin is the presence of cubic couplings of the form $\Phi_i^\dagger 
\Phi_j H$ where $H$ is a heavy scalar field.  If this coupling strength is 
of order $m_H$, then there is a nondecoupling contribution to the low-energy 
$|\Phi_i^\dagger \Phi_j|^2$ interaction.  This does not occur in a 
supersymmetric theory, because this contribution is exactly canceled by 
an existing $|\Phi_i^\dagger \Phi_j|^2$ coupling, or the cubic coupling 
strength itself is much smaller than $m_H$.

We show in particular the nondecoupling of the quartic self-coupling of the 
standard-model Higgs doublet from superheavy scalar bosons in a 
nonsupersymmetric $SU(5)$ grand unified theory.  This has the important 
implication that the usual renormalization-group analysis of the evolution 
of this coupling from the $SU(5)$ scale to the electroweak scale is not valid. 
On the other hand, in a supersymmetric field theory, there is no such problem. 
This is easily understood because supersymmetry relates quartic scalar 
couplings to gauge and Yukawa couplings.  Since the latter do not have 
nondecoupling contributions, neither must the former.

Turning the argument around, we emphasize the possibility that if the 
scale of soft supersymmetry breaking is a few TeV and there exists a 
gauge symmetry larger than that of the standard model just above it, then 
nondecoupling of the TeV-scale physics from the elctroweak Higgs sector 
may occur.  This has been shown explicitly in several previous 
examples\cite{1,2,3}.  It has the important implication that if two 
Higgs doublets are found at the electroweak scale and they are not those 
of the minimal supersymmetric standard model (MSSM), it does not rule out the 
existence of supersymmetry.  There is a large class of supersymmetric 
theories with an extended gauge symmetry at the TeV scale, which has two 
Higgs doublets at the electroweak scale different from those of the MSSM. 
We add one more example here in this paper involving four Higgs doublets. 
Finally we should remark that our results are of course based on 
perturbation theory, but there can also be nonperturbative nondecoupling 
effects through instantons in certain theories\cite{10}.
\vspace{0.3in}
\begin{center} {ACKNOWLEDGEMENT}
\end{center}

This work was supported in part by the U. S. Department of Energy under 
Grant No. DE-FG03-94ER40837.  X.~L. acknowledges the hospitality of the 
Department of Physics, University of California, Riverside during a 
four-week visit when this work was initiated.
\vspace{0.3in}

\bibliographystyle{unsrt}

\end{document}